\documentstyle[multicol,aps,prl]{revtex}

\renewcommand{\narrowtext}{\begin{multicols}{2} \global\columnwidth20.5pc}
\renewcommand{\widetext}{\end{multicols} \global\columnwidth42.5pc}
\def\al{\alpha}
\def\be{\beta}

\def\ze{\zeta}
\def\et{\eta}

\def\la{\lambda}

\def\om{\omega}

\def\De{\Delta}

\def\Om{\Omega}

\def\fr#1#2{{{#1} \over {#2}}}

\def\ket#1{|{#1}\rangle}

\def\frac#1#2{{\textstyle{{#1}\over {#2}}}}

\def\lsim{\mathrel{\rlap{\lower4pt\hbox{\hskip1pt$\sim$}}
    \raise1pt\hbox{$<$}}}
\def\gsim{\mathrel{\rlap{\lower4pt\hbox{\hskip1pt$\sim$}}
    \raise1pt\hbox{$>$}}}
\def\sqr#1#2{{\vcenter{\vbox{\hrule height.#2pt
         \hbox{\vrule width.#2pt height#1pt \kern#1pt
         \vrule width.#2pt}
         \hrule height.#2pt}}}}

\def\lrpartial{\raise 1pt\hbox{$\stackrel\leftrightarrow\partial$}}

\def\aw{$a_\mu^w$}
\def\bw{$b_\mu^w$}
\def\cw{$c_{\mu\nu}^w$}
\def\dw{$d_{\mu\nu}^w$}
\def\ew{$e_\mu^w$}
\def\fw{$f_\mu^w$}
\def\gw{$g_{\la\mu\nu}^w$}
\def\Hw{$H_{\mu\nu}^w$}

\def\tib{\tilde{b}}
\def\tic{\tilde{c}}
\def\tid{\tilde{d}}
\def\tig{\tilde{g}}

\def\X{{\hat X}}
\def\Y{{\hat Y}}
\def\Z{{\hat Z}}

\def\etal{{\it et al.}}

\newcommand{\beq}{\begin{equation}}
\newcommand{\eeq}{\end{equation}}
\newcommand{\bea}{\begin{eqnarray}}
\newcommand{\eea}{\end{eqnarray}}
\newcommand{\rf}[1]{(\ref{#1})}

\begin{document}

\title{Clock-Comparison Tests of Lorentz and CPT Symmetry in Space}

\author{Robert Bluhm,$^a$ V.\ Alan Kosteleck\'y,$^b$
Charles D.\ Lane,$^c$ and Neil Russell$^d$}

\address{$^a$Physics Department, Colby College, 
Waterville, ME 04901}
\address{$^b$Physics Department, Indiana University, 
Bloomington, IN 47405}
\address{$^c$Physics Department, Berry College, 
Mount Berry, GA 30149}
\address{$^d$Physics Department, Northern Michigan University, 
Marquette, MI 49855}

\date{IUHET 440, September 2001}

\maketitle

\begin{abstract}

Clock-comparison experiments conducted in space can provide
access to many unmeasured coefficients for Lorentz and CPT violation.
The orbital configuration of a satellite platform 
and the relatively large velocities attainable in a deep-space mission
would permit a broad range of tests with Planck-scale sensitivity.

\end{abstract}


\narrowtext

A major open challenge in science 
is understanding physics at the Planck scale, 
$m_P\simeq 10^{19}$ GeV.
Direct experimental access to this scale is impractical,
but suppressed effects from it might be observable 
in tests of exceptional sensitivity.
One promising candidate signal is Lorentz violation
\cite{cpt98},
which might arise in string theory 
with or without CPT violation
\cite{kps}
and is a feature of noncommutative field theories
\cite{chklo}.
Observable effects are described by a general standard-model extension
allowing for Lorentz and CPT violation
\cite{ck}.

Among the sharpest tests of Lorentz symmetry in matter
are clock-comparison experiments 
\cite{ccexpt,lh,db,dp}.
These search for spatial anisotropies 
by studying the frequency variation
of a Zeeman hyperfine transition 
as the quantization axis changes orientation.
Traditionally,
the frequencies of two different co-located clocks
are compared as the laboratory rotates
with the Earth.
Experiments of this type 
are sensitive to suppressed effects from the Planck scale
\cite{kla}.
Other tests also constrain various sectors
of the standard-model extension,
involving hadrons 
\cite{kexpt,ak,bexpt,ckpvi,bckp},
photons
\cite{ck,cfj},
muons
\cite{vh},
and electrons 
\cite{eexpt,eexpt2}.

In this work,
we show that clock-comparison experiments
on satellites and other spacecraft 
can provide wide-ranging 
tests of Lorentz and CPT symmetry
with Planck-scale sensitivity.
We consider space experiments in a general theoretical context
and discuss tests for some specific 
orbital and deep-space missions,
including several approved 
for the International Space Station (ISS).

The presence of Lorentz and CPT violation
causes frequency shifts in certain Zeeman hyperfine transitions
\cite{kla}.
In the clock frame,
the relevant contributions to these shifts 
are controlled to leading order by a few parameters
conventionally denoted as
$\tib_3^w$, 
$\tic_q^w$, 
$\tid_3^w$, 
$\tig_d^w$, 
$\tig_q^w$,
where the superscript $w$ is $p$ for the proton,
$n$ for the neutron, and $e$ for the electron.
These parameters are special combinations
of the basic coefficients 
\aw, \bw, \cw, \dw, \ew, \fw, \gw, \Hw\ 
appearing in the standard-model extension
and related to expectation values in the fundamental theory.
For example,
$\tib_3^w = b_3^w -m_w d_{30}^w
+ m_w g_{120}^w -H_{12}^w$,
where $m_w$ is the mass of the particle of type $w$
and the subscripts are indices defined in 
a coordinate system 
with the $3$ direction along the clock quantization axis.

Consider first a clock fixed in a ground-based laboratory.
Then,
the parameters 
$\tib_3^w$, 
$\tic_q^w$, 
$\tid_3^w$, 
$\tig_d^w$, 
$\tig_q^w$
vary in time with periodicities determined by 
the Earth's sidereal angular frequency
$\Om \simeq 2\pi/(23 \, {\rm h} \, 56 \, {\rm min})$.
To display this time dependence,
it is useful to convert these parameters
from the clock frame with coordinates $(0,1,2,3)$
to a nonrotating frame with coordinates $(T,X,Y,Z)$.
The nonrotating frame should for practical purposes 
be an inertial reference frame,
or at least a frame that is inertial
to a degree appropriate to the 
experimental sensitivity.
Possible choices of frame might include,
for example,
ones associated with the Earth, the Sun, the Milky Way galaxy,
or the cosmic microwave background radiation.
Previous literature has restricted
attention to a nonrelativistic conversion 
from the clock frame to the nonrotating frame.
Under these circumstances,
all the above frames are acceptable
and existing experimental bounds are unaffected 
by the choice among them.
However,
in the context of space-based experiments,
the high velocities attainable
make it of interest to consider also
leading-order relativistic effects due to clock boosts.
An Earth-centered choice 
is then no longer appropriate because
it yields distinguishable inertial frames
at different times of year.
In contrast,
frames centered on the Sun, the galaxy,
and the microwave background
each remain unchanged approximate inertial frames
over thousands of years.
Any one of these can be used,
but the choice must be specified when
reporting bounds.

In the experimental context
a Sun-based frame is natural,
and we adopt it here.
For convenience, 
we fix the spatial origin at the Sun's center
with the unit vector $\Z$ along the Earth's rotation axis,
$\X$, $\Y$ in the equatorial plane,
and $\X$ pointing towards the vernal equinox
on the celestial sphere.
The time $T$ is measured by a clock at rest 
at the origin,
with $T=0$ taken as the vernal equinox in the year 2000.
In this frame,
the Earth's orbital plane lies at an angle 
$\et \simeq 23^\circ$ with respect to the $XY$ plane.

For the analysis of space-based experiments,
it suffices to approximate the Earth's orbit as circular
with mean angular frequency $\Om_\oplus$
and mean speed $\be_\oplus$.
Similarly,
a satellite orbit about the Earth can be approximated as circular
with mean angular frequency $\om_s$
and mean speed $\be_s$.
We denote by $\ze$
the angle between $\Z$ and the axis of the satellite orbit 
and by $\al$ the azimuthal angle 
at which the orbital plane intersects 
the Earth's equatorial plane.
Various perturbations cause $\al$ to precess.

In the Sun-based frame,
the instantaneous clock boost 
is $\vec V(T)=d\vec X/dT$,
where the instantaneous spatial location 
$\vec X(T)$ of the clock
depends on the spacecraft and Earth trajectories.
Infinitesimal time intervals in the clock frame
are dilated relative to ones in the Sun-based frame
by an amount controlled by $\vec V(T)$.
An accurate conversion between the two times
must allow for effects such as 
small perturbations in $\vec V(T)$ 
and the gravitational potential.
However,
these complications are irrelevant 
when two clocks at a given location are compared:
conventional relativity predicts an identical rate of advance.
In contrast,
in the presence of Lorentz and CPT violation
two co-located clocks involving different atomic species
typically behave differently,
producing a signal that cannot be mimicked in conventional relativity.

The orientation of the clock quantization axis 
may change as a satellite orbits,
depending on the flight mode.
For brevity in specific examples below,
we assume a flight mode and clock configuration
such that the clock quantization axis 
is instantaneously tangential 
to the satellite's circular trajectory about the Earth.
The clock frame can then be chosen to have
3 axis parallel to the satellite motion about the Earth,
1 axis pointing towards the center of the Earth,
and 2 axis perpendicular to the satellite orbital plane. 
This configuration is,
for example, 
currently planned for some clock experiments aboard the ISS.
However,
our general methodology and results hold  
for arbitrary orientations of the clock quantization axis 
\cite{fn1}
and for various spacecraft flight modes. 

The conversion of a signal in the clock frame
to the Sun-based frame
involves combining the boost $\vec V(T)$ 
with the rotation of the clock as it orbits the Earth.
According to the above discussion,
components of the coefficients for Lorentz violation
in the clock frame
are to be expressed in terms of components 
in the Sun-based frame.
For example,
the component $b_3^w$ becomes
\bea
b_3^w &=& b_T^w \{\be_s - \be_\oplus [\sin \Om_\oplus T
(\cos \al \sin \om_s \De T
  \nonumber \\
&&
\qquad
\qquad
+ \cos \ze \sin \al \cos \om_s \De T )
- \cos \et \cos \Om_\oplus T
  \nonumber \\
&&
\qquad
\qquad
\times (\sin \al \sin \om_s \De T
- \cos \ze \cos \al \cos \om_s \De T )
  \nonumber \\
&&
\qquad
\qquad
+ \sin \et \cos \Om_\oplus T \sin \ze \cos \om_s \De T]\}
  \nonumber \\
&&
- b_X^w (\cos \al \sin \om_s \De T
+ \cos \ze \sin \al \cos \om_s \De T )
  \nonumber \\
&&
- b_Y^w (\sin \al \sin \om_s \De T
- \cos \ze \cos \al \cos \om_s \De T )
  \nonumber \\
&&
+b_Z^w \sin \ze \cos \om_s \De T,
\label{b3}
\eea
where $\De T = T - T_0$ is the time measured from a 
reference time $T_0$.
This equation holds to leading order in linear velocities
and so neglects effects such as the Thomas precession.
The result \rf{b3} for the component $b_3^w$ 
must be combined with results for other coefficients 
to yield the Sun-frame expression 
for the observable parameter $\tib_3^w$.
A similar procedure yields the other observables 
$\tic_q^w$,
$\tid_3^w$,
$\tig_d^w$, 
$\tig_q^w$.
The full expressions are lengthy
and depend on various combinations of
basic coefficients for Lorentz and CPT violation,
on trigonometric functions of various angles
and frequency-time products,
and on $\be_\oplus$ and $\be_s$.

An immediate advantage of space-based experiments
is the direct accessibility of all spatial components 
of the basic coefficients for Lorentz and CPT violation.
Existing ground-based clock-comparison experiments 
seek frequency variations as the Earth rotates,
and the fixed rotational axis implies that the signal
is independent of certain spatial components.
For example,
in these experiments 
the parameter $\tib_3^w$ provides
sensitivity only to the nonrotating-frame components
$\tib_X^w$, $\tib_Y^w$,
which in turn involve a restricted subset of components of
\bw, \dw, \gw, \Hw.
In contrast,
an orbiting satellite can access all spatial components.
Typically,
the satellite orbital plane differs from the
equatorial plane, thus offering different sensitivity
from traditional Earth-based experiments. 
In addition,
the precession of the satellite orbital plane
makes it feasible to sample all spatial directions.

In space,
the relatively short orbital periods
($\om_s \gg \Om$)
imply that the time required for collecting an adequate dataset
can be much reduced.
For example, 
the ISS period is about 92 min,
so an experiment on the ISS
could be completed about 16 times faster
than a traditional Earth-based one,
better matching clock stabilities
and reducing the needed time from months to days.
This makes practical 
an analysis of the leading relativistic effects
due to the instantaneous speed
$\be_\oplus \simeq 1\times 10^{-4}$
of the Earth in the Sun-based frame,
which in turn provides sensitivity to many more
types of Lorentz and CPT violation.
Existing ground-based experiments 
typically take data over months,
during which the Earth's velocity vector changes significantly.
In space,
the shorter timescale for dataset collection 
means that this vector could be treated
as approximately constant.
An experiment could therefore be viewed as 
involving a single inertial frame,
which would allow direct extraction 
of leading relativistic effects.

For space-based experiments,
the above effects combine to yield an overall sensitivity
to many types of Lorentz and CPT violation
that remain unconstrained to date.
Consider,
for example,
a clock-comparison experiment
sensitive to the observable $\tib_3^w$ for some $w$.
In the Sun-based frame and for each $w$,
this observable is a combination 
of the basic coefficients 
\bw, \dw, \gw, \Hw\
for Lorentz violation,
which include 35 independent observable components
once allowance has been made for the 
effect of field redefinitions.
A traditional ground-based experiment  
is sensitive to 8 of these
\cite{fn2}.
We find that
the same type of experiment mounted on a space platform
would acquire sensitivity to all 35.

For some components,
the Lorentz and CPT reach is 
suppressed by a factor of $\be_\oplus$.
This is the dominant linear boost factor 
in the relativistic corrections.
However,
space-based clock-comparison experiments 
would also be sensitive to first-order relativistic effects 
proportional to $\be_s$.
The corresponding effects in traditional Earth-based experiments
would be impractical to study and in any case
would be further suppressed by a factor of $\Om/\om_s$,
which is,
for example,
about $6 \times 10^{-2}$ for the ISS.

Among the order-$\be_s$ effects 
is a seemingly counterintuitive one:
in space-based experiments
a dipole shift can generate a detectable signal 
at frequency $2\om_s$.
This contrasts with the usual analysis of 
ground-based experiments,
where signals with frequency $2\Om$ arise only from quadrupole shifts.
To gain insight about this effect,
consider the parameter $\tib_3^w$.
Nonrelativistically,
this parameter is the third component of a vector
and hence would lead only to a signal at frequency $\om_s$.
However,
$\tib_3^w$ contains $d_{03}$, 
which behaves like a two-tensor
in a relativistic treatment incorporating first-order effects
from $\be_s$
and hence can generate a signal at frequency $2\om_s$. 
Thus,
for example,
when the Earth is near the northern-summer solstice,
the coefficient $C_2$ of $\be_s \cos 2 \om_s \De T$
in the expression for $\tib_3^w$
in the Sun-based frame includes a dependence on 
purely spatial components of \dw: 
\bea
C_2 &\supset& 
\fr m 8 [ \cos 2 \al (3+\cos 2 \ze) (d^w_{XX} - d^w_{YY})
\nonumber \\
&&
+ (1- \cos 2 \ze) (d^w_{XX} + d^w_{YY} - 2 d^w_{ZZ}) 
\nonumber \\
&&
- 2 \sin 2 \ze (\cos \al \, (d^w_{YZ} + d^w_{ZY})
- \sin \al \, (d^w_{ZX} + d^w_{XZ})) 
\nonumber \\
&&
+ (3+\cos 2 \ze)\sin 2 \al \, (d^w_{XY} + d^w_{YX})] .
\label{2oms}
\eea
Monitoring the frequency $2\om_s$ 
would therefore provide sensitivity
to all observable spatial components of \dw.

We focus next on the special case where
the orbiting platform is the ISS,
for which
$\be_s \simeq 3 \times 10^{-5}$
and $\ze \simeq 52^\circ$.
Among the instruments planned for flight on the ISS
are H masers, laser-cooled Cs and Rb clocks,
and superconducting microwave cavity oscillators
\cite{parcs,aces,race,sumo}.
We provide here a simplified theoretical analysis,
applicable to possible Lorentz tests 
with all except the oscillators,
which are discussed elsewhere
\cite{km}.
Note that the practical implementation of these experiments 
would require careful consideration of various technical issues,
including the limitations imposed by the ambient magnetic fields 
on the ISS.
For simplicity,
we assume the signal clock is referenced to a co-located clock
that is insensitive to leading-order Lorentz and CPT violation,
such as an H maser 
operating on its clock transition
$\ket{1,0} \rightarrow \ket{0,0}$
\cite{rv}.

An H maser could also be used as the signal clock.
An experiment could be envisaged
analogous to a recent Earth-based Lorentz and CPT test,
which measured the maser transition 
$\ket{1,\pm 1} \rightarrow \ket{1,0}$
using a double-resonance technique 
\cite{dp}.
This would provide sensitivity to the parameters
$\tib_3^p$ and $\tib_3^e$
in the clock frame
without the interpretational issues
associated with experiments using atoms with more complex nuclei.
The relatively short ISS orbital period implies
only about a day of continuous operation could suffice 
to obtain a dataset roughly comparable to that
obtained over the course of four months 
in a traditional Earth-based experiment.
The orbital inclination ($\ze\neq 0$)
and the possibility of repeating the experiment 
for a different value of $\al$
means that for $w=e,p$ all spatial components 
of \bw, $m_w$\dw, $m_w$\gw, \Hw\ could be sampled.
Assuming that the previous sensitivity of about 500 $\mu$Hz
can also be achieved in space,
several components presently unbounded would
be tested at the level of about $10^{-27}$ GeV,
while others would be tested at about $10^{-23}$ GeV.
Searching also for a signal at frequency $2\om_s$
would permit cleaner bounds on some 
spatial components of $m_w$\dw, $m_w$\gw
at the level of about $10^{-23}$ GeV.
In all,
we find that 
about 50 components of coefficients 
for Lorentz violation
that are currently unconstrained could be measured 
with Planck-scale sensitivities.

In a laser-cooled $^{133}$Cs clock,
the standard clock transition
$\ket{4,0} \rightarrow \ket{3,0}$
is insensitive to Lorentz violation
and could therefore be used as a reference.
For the signal,
a Zeeman hyperfine transition such as
$\ket{4,4} \rightarrow \ket{4,3}$
must be measured.
The electronic configuration of $^{133}$Cs 
involves an unpaired electron,
so the sensitivity to electron parameters
is similar to that of the H maser.
The Schmidt nucleon for $^{133}$Cs is a
proton with angular momentum 7/2,
which would provide sensitivity to all clock-frame parameters 
$\tib_3^p$, 
$\tic_q^p$, 
$\tid_3^p$, 
$\tig_d^p$, 
$\tig_q^p$
and would thus yield both dipole and quadrupole shifts.
In particular, 
components of $c_{\mu\nu}^p$ could be tested.
A traditional ground-based experiment using 
the $\ket{4,4} \rightarrow \ket{4,3}$ transition 
has reached the level of about 50 $\mu$Hz
\cite{lh}.
The duration of an analogous experiment on the ISS would be 
reduced 16-fold.
In addition,
studies of the signal at frequency $2\om_s$
would allow a measurement
of the spatial components of $c_{\mu\nu}^p$ 
at the level of $10^{-25}$
and other components at about $10^{-21}$. 
In this case,
first measurements with Planck-scale sensitivity
of about 60 components of coefficients 
for Lorentz and CPT violation would be possible.

The features of an experiment using $^{87}$Rb 
are similar in many respects.
The standard 
$\ket{2,0} \rightarrow \ket{1,0}$
clock transition 
is insensitive to Lorentz and CPT violation.
However,
a Zeeman hyperfine transition such as
$\ket{2,1} \rightarrow \ket{2,0}$
could be adopted as a signal clock.
Since $^{87}$Rb has an unpaired electron,
its sensitivity to electron parameters
is similar to that of an H maser or 
a Zeeman hyperfine transition in $^{133}$Cs. 
The Schmidt nucleon for $^{87}$Rb is a
proton with angular momentum 3/2,
so the sensitivity to proton parameters is
also analogous to that of the $^{133}$Cs case
up to factors of order unity. 
One potential advantage is that 
the nuclear configuration has magic neutron number,
so theoretical calculations are likely to be more reliable
and experimental results would be cleaner
\cite{kla}.
Like the $^{133}$Cs case,
numerous Lorentz and CPT tests 
with Planck-scale sensitivity
could be performed. 

Other types of spacecraft could also provide
valuable Lorentz and CPT tests.
Speeds an order of magnitude greater 
than $\be_\oplus$ could be accessible
in certain missions.
For example,
the proposed SpaceTime experiment
\cite{spacetime}
would fly co-located 
$^{111}$Cd$^+$,
$^{199}$Hg$^+$,
and $^{171}$Yb$^+$ ion clocks
on a solar-infall trajectory from Jupiter,
attaining $\be \simeq 10^{-3}$.
The craft would rotate several times per minute,
so even 15 min might suffice to acquire 
a complete dataset for Lorentz and CPT tests.
For all three clocks,
the standard clock transitions 
$\ket{1,0} \rightarrow \ket{0,0}$
are insensitive to Lorentz and CPT violation
and so could be used as references.
A signal clock would require monitoring 
a Zeeman hyperfine transition such as
$\ket{1,1} \rightarrow \ket{1,0}$.
The electronic configuration would then permit
sensitivity to electron parameters.
Also, 
the Schmidt nucleon for all three isotopes is
a neutron with angular momentum 1/2,
so all three clocks would be sensitive to
the neutron parameters $\tib_3^n$, $\tid_3^n$, $\tig_d^n$
in the clock frame.
These parameters would be unconstrained by the ISS experiments
discussed above.
Monitoring the signal 
at the spacecraft rotation frequency $\om_{ST}$ 
and also at $2\om_{ST}$ 
would again permit numerous measurements 
of unconstrained coefficients for Lorentz and CPT violation.
The large boost provides 
experiments of this type an intrinsic order of magnitude
greater sensitivity to Lorentz and CPT violation
than measurements performed
either on the Earth 
or in orbiting satellites.

We thank Kurt Gibble, Larry Hunter, John Prestage, and Ron Walsworth
for valuable comments.
This work is supported in part 
by NASA grant NAG8-1770,
by DOE grant DE-FG02-91ER40661,
and by NSF grants PHY-9801869 and PHY-0097982.

\end{multicols}

\begin{references}
\bibitem{fn1}
Note that achieving optimal sensitivity to certain components
can require specific clock orientations.
\end{references}
\end{document}